\documentstyle[prl,twocolumn,aps,epsf]{revtex}

\begin{document}

\twocolumn[\hsize\textwidth\columnwidth\hsize\csname @twocolumnfalse\endcsname
\draft
\tolerance 500

\title{Strongly correlated hopping and many-body bound states}
\author{Julien Vidal$^{1}$,  Benoit Dou\c{c}ot$^{2,3}$}

\address{$^1$ Groupe de Physique des Solides, CNRS UMR 7588,
Universit\'{e}s Paris 6 et 7,\\
2, place Jussieu, 75251 Paris Cedex 05 France}

\address{$^2$ Laboratoire de Physique de la Mati\`ere Condens\'ee, CNRS UMR 8551, 
\'Ecole Normale Sup\'erieure,\\ 
24, rue Lhomond, 75231 Paris Cedex 05 France}

\address{$^3$ Laboratoire de Physique Th\'{e}orique et
Hautes \'Energies, CNRS UMR 7589, Universit\'{e}s Paris 6 et 7,\\ 
4, place Jussieu, 75252 Paris Cedex 05 France}

\maketitle

\begin{abstract}

We study a system in which the quantum dynamics of electrons depend on the particle density in their
neighborhood. For any on-site repulsive interaction, we show that the exact two-body and three-body
ground states are bound states. We also discuss the finite density case in a
mean-field framework and we show that the system can undergo an unusual transition from an effective
attractive interaction to a repulsive one, when varying the electron density. 
  
\end{abstract}

\pacs{PACS numbers: 71.10.Fd, 71.23.An, 73.20.Jc}

\vskip2pc]

%
%
\section{Introduction}
\label{intro}
%
%
Correlated hopping models have been the subject of many studies
within various contexts. First proposed by M. E. Foglio and L. M. Falicov to decribe mixed valence
solids\cite{Foglio_CH}, they have also been widely used to study the organic
conductors\cite{KSSH_CH,Baeriswyl_CH,Gammel_CH,Montorsi} in order to take into account bond-charge
effect\cite{Campbell_CH}. Finally, these models where the probability of an electron to move depends 
on the particle density, have been proposed to mimick effective attractive interaction between electrons
high-$T_c$ superconductivity\cite{Hirsch_CH1,Hirsch_CH2}) and have provided
rich phase diagrams\cite{Essler_CH1,Essler_CH2,Strack_CH,Arrachea_CH1,Karnaukhov_CH,%
Bedurftig_CH,Aligia_CH,Michielsen_CH1,Michielsen_CH2,DeBoer_CH1,Bariev_CH,Airoldi_CH,%
Quaiser_CH,DeBoer_CH2,Arrachea_CH2,Schadschneider_CH,Alcaraz_CH,Japaridze_CH}. 

In a completely different framework, we have recently described a localization phenomenon induced by the
magnetic field that occurs for special geometries and for special values of the magnetic
flux\cite{Vidal_Cages}. This surprising effect has been experimentally observed in superconducting wire
networks\cite{Abilio_T3}, and in two-dimensional electron gas\cite{Naud_T3}.  We have also
studied the influence of electron-electron interactions on such systems and we have shown that a
Hubbard-like term (on-site repulsion) was able to delocalize two particles initially confined in a given
so-called Aharonov-Bohm cage\cite{Vidal_Cages_big,Vidal_Chapelet}. These results have led us to formulate the
simple toy model presented here in which we introduce this delocalization process directly in the hamiltonian by
imposing that an electron can move only if another electron is in its close neighbourhood. Of course, we do
not claim to capture all the physics of the interacting Aharonov-Bohm cages with this simple
one-dimensional system, but we think that it can help us in understanding the delocalization process
induced by (repulsive) interactions.

This paper is organized in two main parts. In the first one, 
we exactly solve the two-body (section \ref{twobody}) and  three-body (section \ref{threebody}) problems and
we show that the ground state is always a bound state for any strength of the on-site repulsion. Therefore, we
give a simple picture in terms of a graph in the Hilbert space that makes the physical interpretation  clearer.
In the second part (section \ref{finite}), we propose a mean-field like approach for the finite density
case by considering the low-energy excitations above the Fermi sea. We show that in the vicinity of a curve
int its two parameter space, the initial hamiltonian can be mapped onto an effective Hubbard model with an
interacting term that depends on the particle density and that can be either repulsive or attractive.

We consider interacting spin $1/2$ fermions in a one-dimensional chain of linear size $L=Na$ with
periodic boundary conditions\cite{notePBC} described by the following hamiltonian~:
%
%
\begin{eqnarray}
H&=&-t \sum_{ i,\sigma}
(c^\dagger_{i+1,\sigma}\,c_{i,\sigma} + c^\dagger_{i,\sigma}\,c_{i+1,\sigma})
(n_{i,-\sigma} +n_{i+1,-\sigma})
\nonumber \\ 
&&+ U \sum_{i} n_{i,\uparrow} \, n_{i,\downarrow}
\mbox{,}
\label{hamil}
\end{eqnarray}
%
%
where $c^\dagger_{i,\sigma}$ (resp. $c_{i,\sigma}$) denotes the creation (resp. annihilation) operator
of a fermion with spin $\sigma$, $n_{i,\sigma}=c^\dagger_{i,\sigma}\,c_{i,\sigma}$ the 
density of spin $\sigma=\uparrow, \downarrow$ fermion on site $i$, and $\langle \ldots \rangle$ stands for
nearest neighbor pairs. The kinetic part of the  hamiltonian (\ref{hamil}) allows a particle of spin
$\sigma$ located on a site
$i$ to jump on a neighbouring site $j$ only if there is already a particle either on site $i$ or on
site $j$. For simpli\-city, we restrict our analysis to the repulsive case $U>0$. However, since the
structure is bipartite, the spectrum of $H$ is odd under the tranformation
$U\rightarrow -U$\cite{Sutherland_Bethe}. 

%
%
\section{The two-body problem}
\label{twobody}
\begin{figure}
\centerline{\epsfxsize=85mm
\epsffile{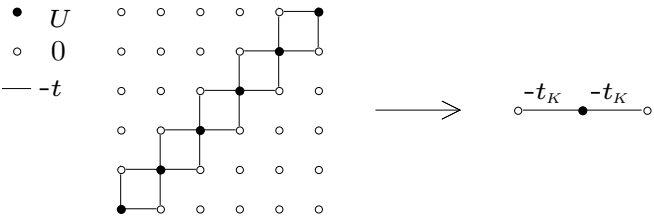}}
\vspace{3mm}
\caption{Representation of the two-body problem as a graph standing in the ${\bf Z}^2$ lattice and
its reduction after projection along (1,1). $t_K=t(1+e^{-iK b})$.}
\label{2-body}
\end{figure}

%
%
The single-body problem is trivially solvable for this model since the particle can neither move, nor
interact. In this case, the spectrum consists in one eigenvalue \mbox{$\varepsilon=0$} which is $N$-fold
degenerate. So, let us pay attention to the two-body problem. Denoting $|i \rangle$ the orbital
localized on the site $i$, the two-body state space is generated by~: 
%
%
\begin{equation}
|i , j \rangle=
|i \rangle_\uparrow \otimes 
|j \rangle_\downarrow, \:\:\: \forall (i,j) \in [0, N-1]^2
\mbox{.}
\label{tensor2}
\end{equation}
%
%
We only consider here the orbital degrees of freedom since for two particles, the spin degrees of freedom
are completely determined by the symmetry of the orbital wavefunction.  
This problem can be mapped on that of a single particle moving in the graph displayed in
Fig.~\ref{2-body} where each site $(i,j)$ of the
${\bf Z}^2$ lattice (with appropriate boundary conditions) corresponds to the ket 
$|i , j\rangle$.  
Using the invariance of the system under a translation of the center of mass 
(direction (1,1) in ${\bf Z}^2$),  diagonalizing $H$ can be simply achieved by introducing the following 
Bloch waves~:
%
%
\begin{eqnarray}
|\varphi_{0}(K) \rangle&=& {1 \over \sqrt{N}} \sum_{n=0}^{N-1}
e^{iK n b} |n, n \rangle\\
|\varphi_{±\pm}(K) \rangle&=& {1 \over \sqrt{N}} \sum_{n=0}^{N-1}
e^{iK n b} |n,n \pm 1 \rangle
\mbox{,}
\end{eqnarray}
%
%
where $b= a \sqrt{2}$  and where $K=2\pi j / N b \:\:\: (j\in [0, N-1])$ is the total momentum of the two
particles. 
%
%
%
%
Indeed, one has~: 
%
%
\begin{eqnarray}
\langle \varphi_{0}(K) |H|\varphi_{\pm}(K') \rangle
&=& -t(1+e^{-iK b}) \delta_{K,K'}\\
&& \nonumber\\
\langle \varphi_{0}(K) |H|\varphi_{0}(K') \rangle
&=& U \delta_{K,K'}
\mbox{,}
\label{Bloch2D}
\end{eqnarray}
%
%
where $\delta_{K,K'}$ is the usual Kronecker symbol, so that the eigenvalues of $H$ are given by
$\varepsilon=0$ for the triplet states (non sensitive to $U$) and for all the trivial configurations
corresponding to isolated particles, and by~:
%
%
\begin{equation}
\varepsilon_{\pm}(K) = {1 \over 2} \left(U \pm \sqrt{U^2+32 t^2 \cos^2(K b /2 )}\right) 
\label{disp2e}
\mbox{,}
\end{equation}
%
%
for the singlet states. Note that this result has already been obtained by Hirsch in a different
context\cite{Hirsch_CH1bis,Hirsch_CH2bis}.  The most surprising fact is the emergence of a dispersive
band
$\varepsilon_{-}(K) <0$ associated to two-body bound states for any $U$. In particular, 
the ground state is obtained for $\varepsilon_{-}(K=0)$. This can be
understood invoking the competition between the kinetic term that lowers the energy and that is enhanced
when  particles are close together, and the interaction term that favors the opposite situation.  
An interesting issue is to know whether this feature still holds for more than two particles. To tackle
this task, we shall now analyze the three-body problem that is, fortunately, still tractable
ana\-lytically.
%
%
\section{The three-body problem}
\label{threebody}
\begin{figure}
\centerline{\epsfxsize=85mm
\epsffile{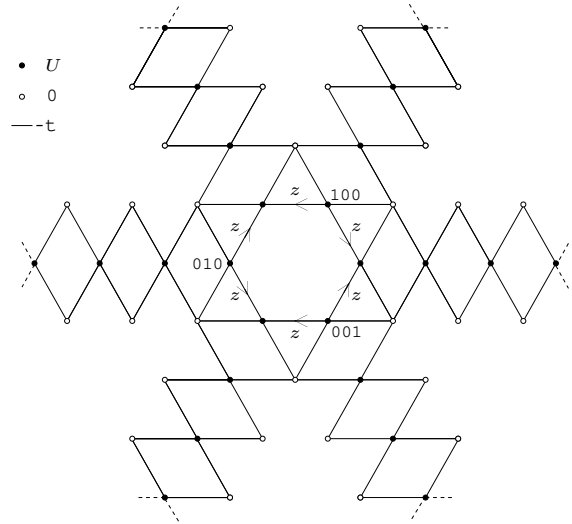}}
\vspace{3mm}
\caption{Representation of the three-body problem graph after projection perpendicularly to the direction
(1,1,1) in the ${\bf Z}^3$ lattice. The sites denoted $i j k$ correspond to the ket $|\varphi_{i,j,k}(K)
\rangle$  and $z=- 2 t e^{-i K c}$.}
\label{sextopussy}
\end{figure}

\begin{figure}[h]
\vspace{-2mm}
\centerline{\epsfxsize=95mm
\epsffile{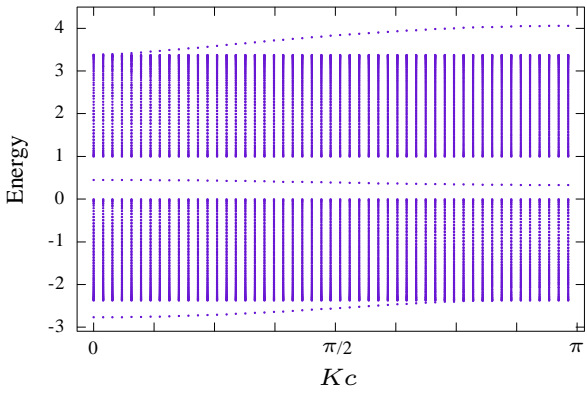}}
\vspace{-6mm}
\caption{Spectrum of the three-body problem as a function of $K c$ for $U=1$ and $N=100$ ($t=1$) .}
\label{dispersion3e}
\end{figure}

%
%
As previously, we consider the three-body problem in the one-dimensional chain as a single-body problem 
in a graph standing in the ${\bf Z}^3$ lattice. Using, once again, the invariance of the system under a
translation of the center of mass (direction (1,1,1) in ${\bf Z}^3$), it is convenient to build the
following Bloch waves~:
%
%
\begin{equation}
|\varphi_{i,j,k}(K) \rangle= {1 \over \sqrt{N}} \sum_{n=0}^{N-1} e^{iK n c} |i+n,j+n,k+n \rangle
\mbox{,}
\label{Bloch3D}
\end{equation}
%
%
where $c= a \sqrt{3}$  and where \mbox{$K=2\pi j / N c \:\:\: (j\in [0, N-1])$} is the total momentum 
of the three fermions. Remark that $|\varphi_{i,j,k}\rangle\propto|\varphi_{i',j',k'}\rangle$ if 
$(i,j,k) \equiv (i',j',k') \: {\em mod} \: (1,1,1)$ so that one must only consider non equivalent Bloch
function. Here, we have chosen to affect the phase $1$ to the ket $|i,j,k \rangle$ such that ($i^2+j^2+k^2$) 
is minimum. After projection along (1,1,1) the graph is the so-called {\it sextopus} displayed in
Fig.~\ref{sextopussy}.
%
%
%
%
Note that in this latter graph, we have not put the site corresponding to $|\varphi_{0,0,0}(K) \rangle$
since it is forbidden for spin $1/2$ fermions but that would be allowed for bosons. 
It is readily seen with this representation that there are two types
of regions~: ({\it i}) the ``bulk" where the three particles are close together and where there could 
possibly exist three-body bound states~; ({\it ii}) the ``legs" corresponding to a situation where one
particle is motionless and the two others propagate together. One also observes that $K$ plays the role
of a  ``pseudo-magnetic field" for the system so that the ground states will be obtained for $K=0$ (zero
field condition).  
For an arbitrary value of $K\neq 0 \:{\em mod} \: \pi / c$, the  {\it sextopus} has the dihedral symmetry $D_3$
and can be easily diagonalized for any $N$. Note that when the three electrons have the same polarization,
the system is frozen since the particles cannot move. It implies that the energy of the quadruplet $S=3/2$ 
is simply 0. The only interesting situation thus arises in the sector $S=1/2$ whose   spectrum is shown in
Fig.~\ref{dispersion3e} as a function of $K$ for a given $U$. 
%
%
%
%

\begin{figure}
\centerline{\epsfxsize=90mm
\epsffile{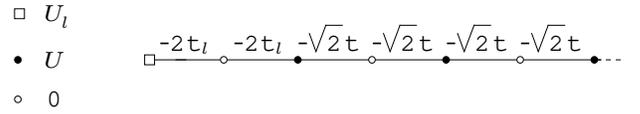}}
\caption{Non trivial graph of each irreducible representation of $D_6$ 
indexed by $l\in[0,5]$. $U_l$ and $t_l$ are defined in the text.}
\label{representation}
\end{figure}

\begin{figure}[h]
\vspace{-2mm}
\centerline{\epsfxsize=95mm
\epsffile{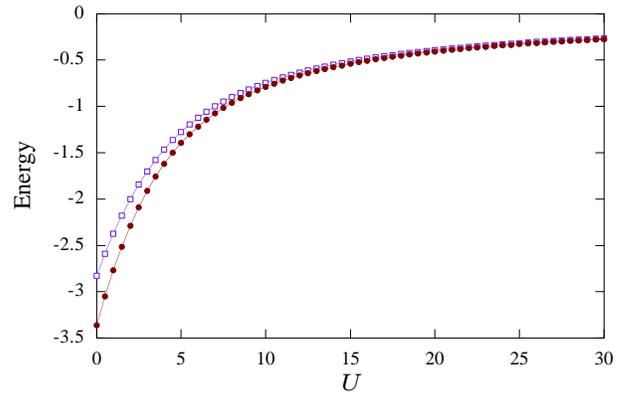}}
\vspace{-6mm}
\caption{Variation of the ground state energy of the two-body $(\Box)$ and three-body $(\bullet)$ problems 
as a function of the interaction $U$ ($t=1$).}
\label{ground}
\end{figure}

One clearly observes two distinct components. First, for a
fixed $K$, there are two dispersive bands that correponds to scattering states (one two-body bound
state + one motionless particle) propagating ballistically in the ``legs". Their precise
form is given by the equation  (\ref{disp2e}). Second, one observes much more interesting
states out of those bands. As we shall see thereafter, these states are actually three-body
bound states and the most remarkable fact is that the ground state is always  given by such
a state for any $U$. To analyze more precisely this surprising phenomenon, we focus on the
case $K=0$ for which the ground state is obtained. In this case, the symmetry of the {\it
sextopus} is $D_6$ and it is possible to look for bound states in each representation
schematized in Fig.~\ref{representation}.
Each site represented in Fig.~\ref{representation} symbolizes a state vector with fixed angular
momentum. 
Note that the representations indexed by $l=0,3$ corresponds to a fully space symmetric
wavefunction that can, in no case, be an eigenstate for three spin $1/2$ fermions.
%
%
%
%
We thus look for a bound eigenstate $|\Psi\rangle=\sum_n\psi_n |n\rangle$ where $|n\rangle$ is the ket
corresponding to the $n^{th}$ site of the half-chains labelled by $l\in [1,2]$ (or
equivalently $l\in [4,5]$). For all $n\geq 3$, we set $\psi_n=C_\lambda e^{-\lambda 
(n-3) b}$ for $n$ odd and $\psi_n=D_\lambda e^{-\lambda (n-3) b}$ for $n$ even, and we
seek for complex $\lambda$ such that
$\Re (\lambda b) \geq 0$ and $\Im(\lambda b) \equiv 0 \:{\em mod} \: \pi$, this latter condition ensuring to
have a real   eigenenergy. 
Such a state is an eigenstate if the  secular equations are satisfied~:
%
%
\begin{eqnarray}
(\varepsilon_\pm^B(\lambda)-U_l) \:\psi_1&=&-2t_l \:\psi_2 \nonumber \\
\varepsilon_\pm^B(\lambda) \:\psi_2&=&-2t_l \:(\psi_1+\psi_3) 
\label{secular}\\
(\varepsilon_\pm^B(\lambda) -U) \:\psi_3&=&-2t_l \:\psi_2 - \sqrt{2}t \:\psi_4 \nonumber
\mbox{,}
\end{eqnarray}
where $U_l=U- 4 t\cos(2\pi l/6)$ and $t_l=t \cos(\pi l/6)$ and~:
%
%
\begin{equation}
\varepsilon_\pm^B (\lambda) = {1 \over 2} \left(U \pm \sqrt{U^2+32 t^2 \cosh^2(\lambda b/2)}\right) 
\label{bound}
\mbox{.}
\end{equation}
%
%
We emphasize that the periodic boundary condition adds restrictions on $\lambda$ but since the
amplitude decreases exponentially, they are not relevant provided $N$ is larger than the
localization length
$1/ \lambda$. 
A simple inspection of the secular system (\ref{secular}) allows to show that it always has, at
least, one solution $\lambda_0$ (for any $U>0$) associated to  $\varepsilon_-^B (\lambda_0)$ which is, 
thereby,  the ground state of the system. 
We have plotted in Fig.~\ref{ground} the dependence of the two-body and three-body 
ground states energy as a function of $U$ for spin 1/2 fermions. 
%
%
%
%
When $U$ goes to infinity, both energies converges toward zero but, as noticed
previously, the three-body ground state has always a lower energy than the two-body one. In addition, the
localization length $1/ \lambda_0$ tends to infinity which sounds quite natural for strong repulsion.

%
%
\section{The finite density case}
\label{finite}
\begin{figure}[h]
\vspace{-2mm}
\centerline{\epsfxsize=50mm
\epsffile{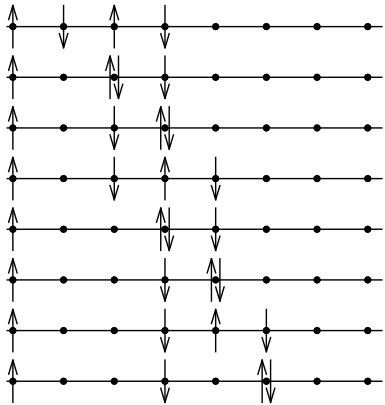}}
\vspace{3mm}
\caption{A possible way to obtain an isolated pair and two frozen electrons from a four-body compact
configuration.}
\label{sequence}
\end{figure}

%
%
A natural question then arises~: are there still many-body bound states at higher densities~? 
Indeed, if we now consider four electrons, we can figure out more complex scattering processes where two
two-body bound states collide with each other, or where a bound pair oscillates between two
isolated particles. In these cases, one may expect the emergence of four-body bound states. 
Unfortunately, this problem is too complicated to be completely analyzed by elementary methods. We have
searched for the possibility of binding together two two-body ground states by means of a variational
method. In this approach, the four-particle state space has been decomposed in two sectors. The first one
corresponds to two isolated two-particle states whose centers of mass  are separated by at least 2.5 lattice
spacings. In this region, we assumed an exponential decay of the wavefunction with the relative separation
of the two center of mass positions. The second region corresponds to real space configurations where each of
the four particles is close to any other.  In the simplest version, this defines, after taking global
translation invariance into account, a finite set of 18 configurations. It is important to note that two
two-body bound states collision may generate real space configurations where a bound pair escapes to
infinity leaving two unpaired electrons which may remain at an arbitrary large final distance from each
other. This is illustrated on the sequence shown in Fig. \ref{sequence}. 
%
%
%
%
However, the typical decay length for the escape of a single pair is likely to be significantly shorter than the
one for two bound pairs. Thus, we have not kept these asymptotic states with a single pair in our variational
approach. This search for four-particle bound states has always failed so far.  Of course, this lack of
evidence does not constitute a proof of the absence of such bound states in the spectrum, but we believe that
four-particle binding is unlikely in this model. The main reason for that is the Pauli principle which severely
restricts the possibility of particle hopping in a close packing configuration. Therefore, we suggest that the
finite but small density system will form a Luttinger liquid of bound pairs. The absence of four-particle
binding may mean that residual interactions between these bound pairs are repulsive, allowing a well-defined
thermodynamic limit. This picture of a fluid of bound pairs is in fact reminiscent of the attractive Hubbard
model\cite{Sutherland_Bethe}. Indeed, a detailed analysis of the Bethe ansatz spectrum\cite{Woynarovich}
shows that this model is precisely described in its low-energy limit in terms of a Luttinger liquid of
spinless two-particle bound states. By analogy, we thus also expect a gap in the spin excitation
spectrum of our model. Actually, the ground state and the thermodynamical properties of both models
should be quite similar. However, dynamical quantities might produce some meaningful differences. For
instance, it would be very interesting to investigate in more details the behaviour of the single
electron Green's function. In our simple-minded picture, adding an  electron to a state with an even
number of particles indeed leaves an unpaired spin. By contrast to the usual spin charge separated
liquid and to the attractive Hubbard model in particular, our model opens the possibility of sustaining
three-body bound states between this extra electron and a pair taken from the Luttinger liquid ground
state. This interesting question clearly deserves further studies. Instead of adressing this issue
here, we shall turn to a very simple description of the finite density system. For large particle
densities, the two-particle bound states have to overlap in real space, and it is no longer clear that
they provide a good basis to understand the ground state properties. In the following, we shall assume
that the quantum state of the finite density system is not too far from a Slater determinant of plane
waves. Let us now write the hamiltonian (\ref{hamil}) in momentum space~:
%
%
\begin{eqnarray}
H &=& {1 \over N} \sum_{k,k',p,p',\sigma} \delta_{p+p',k+k'}  \times \nonumber\\
&& f(p,p';k',k) \:c^\dagger_{p,\sigma} c^\dagger_{p',-\sigma} c_{k',-\sigma} c_{k,\sigma}
\label{hamilFourier}
\mbox{,}
\end{eqnarray}
%
%
where $c_{p,\sigma}={1 \over \sqrt{N}} \sum_n e^{i n p a} c_{n,\sigma}$ and $p$ is the momentum of the
state. We keep the same notation $c$ for operators either in real or in momentum space since 
no confusion is possible here. The interaction function is given by~:
%
%
\begin{eqnarray}
 f(p,p';k',k)&=&{U \over 2}-t[\cos(pa)+\cos(p'a)+  \nonumber \\
             & & \cos(k'a)+\cos(ka)]
\mbox{.}
\end{eqnarray}
%
%
The expectation value of $H$ taken on any free particle state written using the plane wave single particle
basis is given by~:
%
%
\begin{equation}
\langle H \rangle= {2\over L} \sum_{k,k'} f(k,k';k',k) \: n_{k,\uparrow} n_{k',\downarrow}
\mbox{.}
\end{equation}
%
%
To determine the most stable Slater determinant of plane waves, we consider a single particle-hole
excitation away from a reference state whose occupation numbers in momentum space are denoted by
$n^{(0)}_{k,\sigma}$. For instance, we consider~:
$n_{k,\uparrow}=n^{(0)}_{k,\uparrow}+\delta_{k,p}-\delta_{k,q}$ and
$n_{k,\downarrow}=n^{(0)}_{k,\downarrow}$. The average energy change induced by this particle-hole
excitation is~:
%
%
\begin{eqnarray}
\langle  \delta H \rangle_{p,q}&=& {2\over L} \sum_{k'} [f(p,k';k',p) - f(q,k';k',q)] 
 n^{(0)}_{k,\downarrow}\\
                               &=& -4 t {N_{\downarrow}\over N}  [\cos(pa)-\cos(qa)]
\mbox{.}
\end{eqnarray}
%
%
where $N_{\sigma}$ is the total number of spin $\sigma$ electrons. This shows that the average
energy is minimal provided the occupied state (momenta)  fills
the usual Fermi intervals $]-k_{F,\sigma},k_{F,\sigma}]$ for $\sigma=\uparrow, \downarrow$. From these
considerations, and from the expression (\ref{hamilFourier}), it is very natural to separate the diagonal
and off-diagonal part of $H$. We thus have~:
%
%
\begin{equation}
H_{\it{diag}}= {U \over N} N_\uparrow N_\downarrow -2 \sum_{k,\sigma} t_{\it{eff},\sigma} \cos(ka)
c^\dagger_{k,\sigma} c_{k,\sigma}
\mbox{.}
\end{equation}
%
%
with $t_{\it{eff},\sigma}=2 t N_{-\sigma}/N$. Up to a global shift, $H_{\it{diag}}$ exhibits
the same structure as a pure hopping hamiltonian for free particles, but its hopping term depends on the
particle density as it is induced by a two-particle interaction process. Along the same line, one has~:
%
%
\begin{eqnarray}
H_{\it{off-diag}}&=& {1 \over  N} \sum_{k,k',p,p',\sigma} \delta_{k+k',p+p'}  \: (1-\delta_{k,p}) \times
\nonumber\\ &&          f(p,p',k',p) \: 
c^\dagger_{p,\sigma} c^\dagger_{p',-\sigma}  c_{k',-\sigma} c_{k,\sigma}
\label{hoff}
\mbox{.}
\end{eqnarray}
%
%
Let us assume that the ground state is not too remote from the usual non-interacting Fermi sea. In this
case, the four wave vectors involved in (\ref{hoff}) are close to Fermi points so that $H_{\it{off-diag}}$
is well approximated by~:
%
%
\begin{eqnarray}
H_{\it{off-diag}}&\simeq& {1 \over  N} \sum_{k,k',p,p',\sigma} \delta_{k+k',p+p'}  \: (1-\delta_{k,p}) \times
\nonumber\\ &&          {U_{\it eff} \over 2} \: 
c^\dagger_{p,\sigma} c^\dagger_{p',-\sigma}  c_{k',-\sigma} c_{k,\sigma}
\label{hoffbis}
\mbox{.}
\end{eqnarray}
%
%
with $U_{\it eff}=U-8t \cos(k_F a)$. This result is similar to the one obtained by Airoldi and
Parola in a more general model\cite{Airoldi_CH}.
Expression (\ref{hoffbis}) is expected to be valid when
$|U_{\it eff}|$ is smaller than the effective bandwidth which is of the order 
$t (N_\uparrow+N_\downarrow)/N$. When this condition is satisfied, we see that the model becomes equivalent
to a one-dimensional Hubbard model with an effective interaction parameter $U_{\it eff}$. Note that the
terms we have neglected going from (\ref{hoff}) to (\ref{hoffbis}) involve the momentum dependence of
the bare interaction vertex on the external legs so they are irrelevant in the usual renormalization group
analysis of weakly inter\-acting fermions in one dimension\cite{Shankar}. The most interest\-ing consequence of 
(\ref{hoffbis}) is obtained for a less than half-filled system ($|k_F a| < \pi/2$) so that $\cos(k_F a) >0$.
In this case, the model exhibits a qualitative change from an effective attraction to an effective
repulsion as the bare on-site repulsion $U$ crosses the value $8t \cos(k_F a)$. From the previous argument,
we expect a one-dimensional Fermi liquid for $U=8t \cos(k_F a)$ as the off-diagonal interaction is then
purely composed of irrelevant terms. As the density increases, the size of the attractive regions is
reduced since the critical value of $U$ decreases. Note that for dilute systems, this effective Hubbard
model is supposed to be valid only for a narrow interval of values of $U$ around $8t \cos(k_F a)$ since the
effective single electron bandwith is small. In the dilute and attractive regime, it is therefore more
accurate to work within the picture of a Bose gas of bound pairs.

\section{Conclusions}

To conclude, we have shown that a simple one-dimensional model in which the particle hopping is completely
assisted by the presence of another particle in its neighborhood generates some interesting conducting
states at finite density. For the less than half-filled system, in the presence of an on-site repulsion
$U$, we have established the existence of two regimes for the effective interaction  depending on the
strength of $U$. In the attractive regime, the elementary building blocks of the system are likely to be
the two-particle bound states discussed in section \ref{twobody}. Indeed, these states appear for any
value of $U$ and they are characterized by a very tight binding in real space. In this regime, our model is
strongly reminiscent of the one-dimensional attractive  Hubbard model. An interesting
difference remains the existence of three-particle  bound states detailed in section \ref{threebody} that
are absent in the attractive Hubbard model\cite{Sutherland_Bethe}. In the present work, the existence of
these bound states have only been established in the zero density limit. It would be interesting to know whether
they could survive in the presence of a finite particle density, since they could eventually show up in the
single electron spectral function.

Finally, we would like to make a few comments regarding the original motivation of this study, namely the
investigation of interaction effects in tight-binding models for which all single-particle eigenstates are
localized in Aharonov-Bohm cages. A quasi-one-dimensional example illustrating this
phenomenon has been analyzed in details in \cite{Vidal_Chapelet} for the two-body problem. Though the
results obtained here have strong similarities with this latter model, they have also important qualitative
differences. A common point is the possibility to form extended states for the two-particle problem which
are tightly bound in real space independently of the local bare repulsion strength. However, the main
difference lies in the fact that for Aharonov-Bohm cage models, these extended two-particle eigenstates are, at
best, degenerate with the localized ground states and they are in most cases excited states
\cite{Vidal_Cages_big}. In the finite density case, it is by no means obvious that these strongly localized
systems become conducting by contrast to the one studied here. Nevertheless, we believe that if such
conducting states appear in the presence of Aharonov-Bohm cages, two-particle bound states will play a
major role. In this context, such models with strongly correlated (or purely assisted) hopping may provide a
simple and efficient way to describe possible conducting states induced by interactions in otherwise
localized systems.

\acknowledgments

We are indebted to A. Parola and J. E. Hirsch for mentioning us their work on correlated hopping model in the
framework of the high-$T_c$ superconductivity.  We would also like to thank Cl. Aslangul, P. Lecheminant, 
R. Mosseri and  D. Mouhanna for fruitful discussions.



\end{document}